\documentclass[aps, prl, twocolumn, superscriptaddress, showpacs]{revtex4}

\usepackage{amsfonts,amssymb,amsmath,amsbsy}
\usepackage{epsfig}

\usepackage{graphicx}
\usepackage{url}

\begin{document}

\title{Learning by message-passing in networks of discrete synapses}

\author{Alfredo Braunstein }

\affiliation{ICTP, Strada Costiera 11, I-34100 Trieste, Italy}

\author{Riccardo Zecchina}

\affiliation{ICTP, Strada Costiera 11, I-34100 Trieste, Italy}

\begin{abstract}
We show that a message-passing process allows to store in binary \char`\"{}material\char`\"{}
synapses a number of random patterns which almost saturates the information
theoretic bounds. We apply the learning algorithm to networks characterized
by a wide range of different connection topologies and of size comparable
with that of biological systems (e.g. $n\simeq10^{5}-10^{6}$). The
algorithm can be turned into an on-line --fault tolerant-- learning
protocol of potential interest in modeling aspects of synaptic plasticity
and in building  neuromorphic devices.
\end{abstract}
\maketitle

Learning and memory are implemented in neural systems mostly through
distributed changes of synaptic efficacy~\cite{Hebb}.
The learning problem in neural networks (NN) asks whether one can
find values for the synaptic efficacies such that a set of $p$ patterns
are stored simultaneously. Depending on the structure of the network
--- feed-forward or recurrent -- the storage problem is either seen
as a classification problem (input patterns are classified according
to the output of the network) or as an attractor dynamics problem
(patterns are the external stimuli which drive the dynamics of the
network to the closest attractor)~\cite{NeuralNets}. In any case,
understanding the mechanisms underlying synaptic changes constitutes a crucial 
step  for modeling real neural circuits (e.g. the Purkinje cells in the cerebellum~\cite{Purkinje}).
On the purely theoretical side many basic results have been derived,
ranging from information theoretic bounds~\cite{InfoBounds,RigorousBinary}
and statistical physics analysis of learning capabilities~\cite{StatMech}
in model NN to concrete algorithms, like artificial pattern recognition
systems. Still there exist many open conceptual problems that are
related to the need of satisfying realistic constraints~\cite{realistic-constraints}.
Modeling material synapses is possibly one of the most basic ones,
the discrete case (and specifically the switch-like binary one) being
of particular experimental~\cite{discrete} and technological interest~\cite{hardware}:
recent experiments -- at the single synapse resolution level -- have
shown that some synapses undergo potentiation or depression between
a restricted number of discrete stable states through switch-like
unitary events~\cite{discrete}. 
It is has been known since many years that the discreteness of synaptic
efficacies makes the learning problem extraordinarily difficult~\cite{complexity}:
even the task of finding binary synaptic weights for
a single layer network (the binary perceptron) which classifies in two classes a given set of patterns is both NP-complete
and computationally hard on average (as observed in classical numerical
experiments). In spite of the fact that binary networks can in principle
classify correctly an extensive number $p=\alpha n$ of random patterns
with $n$ binary synapses~\cite{capacity}, practically there exists
no known algorithm which is able to store exactly more than just a
logarithmic number~\cite{typical-hardness,Amit-Fusi} as soon as a sub-exponential
cut is put on their running time. 

Here we present a distributed \emph{message-passing} algorithm
of statistical physics origin which is able to store efficiently an
extensive number ($p=\alpha n$ with $\alpha>0$) of random patterns
in binary NN characterized by a wide range of different topologies.
We consider single and multi-layer networks with local connectivities
of the neurons ranging from finite to extensive. The typical computational
complexity of the algorithm will be shown to scale roughly as $O\left(n^{2}\log\left(n\right)\right)$,
that is almost \emph{linearly} on the size of the input for an extensive
number of patterns. This fact together with the parallel nature of
the algorithm allows to easily find optimal synaptic weights for systems
as large as $n=10^{6}$ with $\alpha$
relatively close the critical value $\alpha_{c}$ above which perfect
learning is no longer possible. 
From the algorithmic viewpoint, our solution to the binary learning
problem should be seen as an example of solution
of constraint satisfaction problems over dense \textit{factor graphs}
(a graphical representation of combinatorial constraints used in information
theory~\cite{learning-codes-physics,BP-review}). As such, our result
show how the recent progress in combinatorial optimization by statistical
physics and message passing techniques which have allowed to solve
efficiently famous combinatorial problems like random $K$-satisfiability~\cite{mz-pre}
or random graph $Q$-coloring~\cite{coloring}, can be extended to
other classes of problems in which constraints involve an extensive
number of variables. 

The NN models that we shall consider are composed of simple threshold
units connected by binary weights $w_{j,k}=\pm1$. For the sake of
simplicity we consider two layer networks with one output unit and
with weights of the output layer that are fixed $w_{\ell,out}=1$
(see Fig. 1). Each of the $K$ internal units is connected to $c_{\ell}$ inputs
in either a tree-like structure or in an overlapping way. We will
consider NN with connectivities ranging from finite to extensive,
i.e. take $c_{\ell}=O\left(n^{\epsilon}\right)$ where $\epsilon\in[0,1]$.
In order to keep extensive the overall number of synapses we chose
$K\propto\left\langle c\right\rangle ^{-1}$, where $\left\langle c\right\rangle $
is the average connectivity. Under these conditions, the information
theoretic bounds on the maximum number of bits which can be stored
in the binary synapses are compatible with the exact storage of an
extensive number of patterns ($p=\alpha n,\;\;\alpha>0$)~\cite{StatMech}.
The output $\tau_{\ell}$ of each internal unit is just the sign of
the weighted sum of its inputs $\xi_{j}$ minus some threshold, $\tau_{\ell}=\mbox{sign}(\sum_{j\in V\left(\ell\right)}w_{j,\ell}\xi_{j}-\gamma_{\ell})$
where $V(\ell)$ is the set of inputs connected to unit $\ell$. The
overall output $\sigma$ of the network is given by $\sigma(\vec{\xi})=\mbox{sign}(\sum_{k=1}^{K}\tau_{k}-\gamma_{out})$.

For $K=1$ and $c=n$ we recover the binary perceptron, which is the
elementary building block of many NN models. In the case of random
input patterns, statistical mechanics and rigorous methods~\cite{capacity,StatMech,InfoBounds,RigorousBinary}
have allowed to study the typical behaviour of this type of systems
in the limit of large $n$. For instance the storage capacity $\alpha_{c}$
has been computed for different finite values of $K$. Interestingly
enough, the general scenario for binary networks is that while the
storage capacity is indeed extensive the geometric structure of the
space of solutions in the \textit{satisfiable} region $\alpha<\alpha_{c}$
is rather complex~\cite{geometric-structure-binary}. Optimal synaptic
configurations are typically far apart in Hamming distance and coexist
with an exponential number of sub-optimal configurations in which
an extensive number of errors are made. 
Sub-optimal states act as dynamical traps for learning algorithms~\cite{typical-hardness}.
\begin{figure}
\begin{center}\includegraphics[%
  width=0.80\columnwidth]{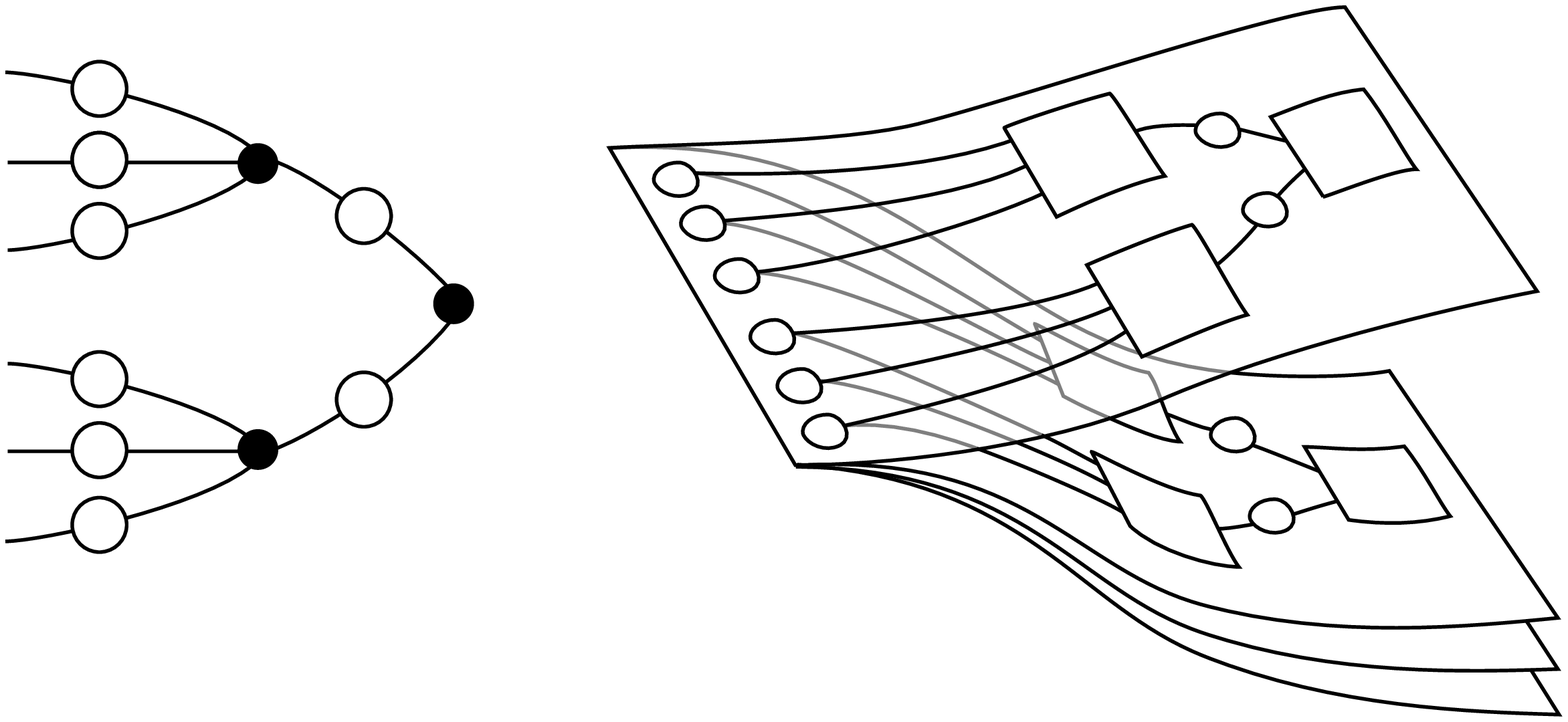}\end{center}
\caption{A non-overlapping two-layer network with six synapses (empty circles)
and three threshold units (filled dots), and its corresponding factor
graph for four patterns (right). The factor graph is composed of \emph{variable}
nodes (circles; indices $i,j,o$ in the text) and \emph{function}
nodes (squares; indices $a,b$ in the text); messages "travel''
over the edges of the factor graph in both directions. Note that while
synaptic weights have a unique corresponding variable node on the
factor graph, each of the two auxiliary variable computing a partial
threshold (\emph{hidden} \emph{units}), being pattern-dependent, must
be replicated for every pattern on the factor graph.\label{fig:graph}}
\end{figure}
Here we first show how the so-called \emph{belief}
\emph{propagation} (BP) equations~\cite{BP-review,learning-codes-physics} 
(a variant of the Bethe approximation in statistical physics) 
can be applied on single problem instances, providing useful information such as
the entropy of solutions, agreeing with statistical physics results in the
large $n$ limit~\cite{capacity}. Next we modify the equations by
introducing a local reinforcement term which forces the system to
polarize to a single optimal configuration of synaptic weights, effectively
turning BP into a solver for this problem.

For simplicity let's fix a threshold value $\gamma$ and first consider
a perceptron with binary weights $w_{i}\in\left\{ -1,1\right\} $
for $i=1,\dots,n$. Given an input pattern $\xi$, the binary perceptron
is an elementary device which just computes the function $f_{\mathbf{w}}\left(\xi\right)=\textrm{sign}\left(\sum_{i}w_{i}\xi_{i}-\gamma\right)\in\left\{ -1,1\right\} $.
Patterns $\xi$ will be then classified by this perceptron by its
output into the two preimage sets of the function $f_{\mathbf{w}}$.
Given two sets of random patterns $\Xi_{\pm}$ we want to find vector
of synaptic weights $\mathbf{w}$ such that $f_{\mathbf{w}}\left(\Xi_{\pm}\right)=\pm1$.
Consider the uniform probability space over the set $W$ of all optimal
assignment. We are interested in single marginals, that is the probabilities
$P\left(w_{i}=\pm1\right)$ that the single synapses take a certain
binary value. Under some weak correlations assumption, it is possible to write
a close set of equations for these quantities. Such BP equations provide results which are believed to be exact in certain classes of problems defined over sparse factor graphs in which the size of
loops tends to infinity with the problem size (e.g. in low density
parity check codes~ \cite{BP-review}).
In the case of problems corresponding to highly connected factor graphs
(like the learning problem we discuss here) the validity of the BP
approach relies on an apparently stronger condition, the so called
\textit{clustering hypothesis}, in which the weak correlations condition
arises from the weak effective interactions among variables. Until
recently no algorithmic approach existed that allowed to study the
properties of a given problem instance of this type. Previous attempts
in this direction were based on iterations of the mean--field TAP
equations~\cite{tap}, which turn out to diverge in most cases. Recently
BP has been used to study some densely connected problems on which
it was shown that BP equations converge while TAP equations do not,
even though the fixed point of the two is the same~\cite{Kabashima}. 

\begin{figure}
\begin{center}\includegraphics[
  width=0.90\columnwidth]{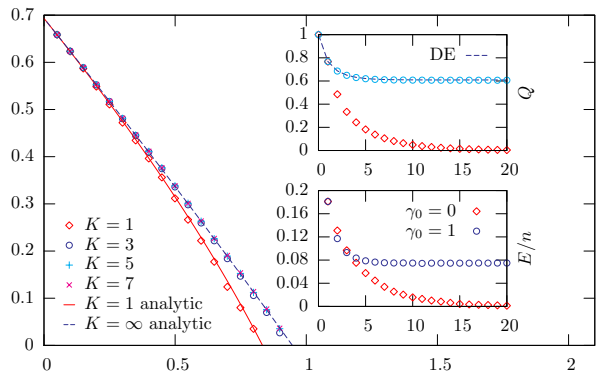}\end{center}
\caption{BP entropy vs. $\alpha$ for single problem instances of size $n=3465$
for $K=1,3,5,7$. The analytic result for $K=1$ and $K\gg1$ for
$n\rightarrow\infty$ are also plotted for comparison. The upper inset
shows $Q^{t}$ vs. $t$ of the analytical DE prediction (dashed line)
vs. simulations over a system of size $10^{5}+1$ at $\alpha=0.6$
without reinforcement (data in perfect agreement to the prediction)
and with reinforcement ($\gamma_{0}=0$). The bottom inset shows the
fraction of errors $E/n$ vs $t$ for both cases. In the latter case
we can see that $Q^{t}\rightarrow0$ as the solution is reached. \label{fig:entropy}}
\end{figure}

\begin{figure}
\centering
\includegraphics[
  width=1.0\columnwidth]{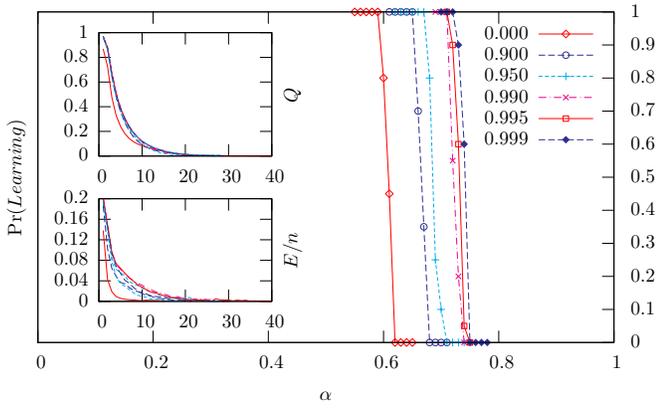}
\caption{Learning of $\alpha n$ pseudo-random patterns curves for the binary
perceptron for different values of $\gamma_{0}$ ($n=10^{4}+1$, 20
samples). The running time scales with $\gamma_{0}$ roughly as $1/(1-\gamma_{0})$.
Inset: evolution of $Q^{t}$and $E^{t}$ vs. time $t$ for various
kinds of two-layer network topologies, i.e. $n=3^{7},\alpha=0.5$
and $K\in\left\{ 3^{0},3^{1},\dots,3^{6}\right\} .$ Note that the
number of errors $E$ goes to $0$ in all cases.\label{fig:learning} }
\end{figure}

At variance with statistical mechanics results where the average
over the patterns and the limit $n\rightarrow\infty$ are done, here
we are interested in single problem instances. Thanks
to the concentration of measure of the error-energy function, the
so called self-averaging property, we expect the quantities estimated by the equations on single problem instances to match the typical case as
$n$ gets large enough. Despite the fact that the approximations behind
BP become exact only as $n$ gets large, also at finite $n$ the results
provide very good approximations which can be used for algorithmic
purposes (see Fig. \ref{fig:entropy}). A large $n$ expansion of
the BP equations for the $K=1$ and $\gamma=0$ network learning problem
read:
\begin{eqnarray}
m_{i\rightarrow a}^{t} & = & \tanh\left(h_{i\rightarrow a}^{t}\right)\label{eq:rules-bp1} \\
u_{b\rightarrow i}^{t} & = & f\left(\frac{1}{\sqrt{n}}\sum_{k\neq i}\xi_{k}^{b}m_{k\rightarrow b}^{t},\frac{1}{n}\sum_{k\neq i}\left(m_{k\rightarrow b}^{t}\right)^{2}\right)\label{eq:rules-bp1.5}\\
h_{i\rightarrow a}^{t+1} & = & \frac{1}{\sqrt{n}}\sum_{b\neq a}\xi_{i}^{b}u_{b\rightarrow i}^{t}\label{eq:rules-bp2}
\end{eqnarray}
 where $f\left(a,b\right)=\left(\int_{0}^{+\infty}\exp\left(-\frac{\left(x-a\right)^{2}-a^{2}}{2\left(1-b\right)}\right)dx\right)^{-1}$.
At the fixed point $m_{i\rightarrow a}$ represents the mean value
of $w_{i}$ over the set of $W^{\left(a\right)}$ of synaptic weight
configurations satisfying all patterns except pattern $\xi^{a}$.
The quantity $h_{i\rightarrow a}$ is referred to as \textit{local
field} that synapse $i$ feels in absence of pattern $a$. The fixed
point of these equations provide the information we are seeking for.
Solving the equations by iteration proved itself to be an efficient
technique, fully distributed, which is known as a \emph{message-passing}
method (the components of the vectors $u$ and $h$ can be thought
as messages running along edges of the factor graph, see Fig.~\ref{fig:graph}).
From the fixed point we may compute the list of all probability marginals
$P\left(w_{i}=\pm1\right)$ together with global quantities of interest
such as the entropy (normalized logarithm of the size of the set $W$).
As expected from the statistical mechanics results~\cite{capacity},
the entropy is monotonically decreasing with $\alpha$ and vanishes
at $\alpha_{c}\sim0.833$ for $n$ large enough. 
Similar results can be derived for multilayer networks as shown in
Fig.~\ref{fig:entropy}. The BP equations can be adapted in a straightforward
way to networks of arbitrary topology, even if the notation is slightly
more encumbered. In general this network will be formed by connecting
several perceptron sub-units. The corresponding factor graph can be
recovered trivially as in Fig.~\ref{fig:graph}, by just replicating
every perceptron for each pattern, and adding a set of auxiliary units
to represent the output of every perceptron sub-unit of the network.
It will suffice then to derive a set of slightly more general BP equations
for the perceptron which we omit for the sake of brevity.
We have studied analytically the dynamical behaviour of the BP algorithm
in the large $n$ limit by the so called \emph{density} \emph{evolution}
(DE) technique (see e.g. \cite{Kabashima} for details on DE). In
the upper inset of Figure~\ref{fig:entropy} we can see the comparison
of numerical simulations of large single instances with the analytical
prediction of the quantity $Q=1-\frac{1}{\alpha n^{2}}\sum_{i}\sum_{a}m_{i\rightarrow a}^{2}$
at every iteration step. 
In the spirit of~\cite{mz-pre}, a way of using the information
provided by BP is to ``decimate'' the problem. This approach
is indeed feasible and leads to optimal assignments. However here
we focus on a much more efficient and fully distributed version~\cite{distributed-SP}
of the algorithm. The idea is to introduce an extra term into 
Eqs.~\ref{eq:rules-bp1}-\ref{eq:rules-bp2} enforcing $h_{i}=\pm\infty$ at a fixed point, and use $w_{i}=\mbox{sign}\left(h_{i}\right)$
as a solution.  This term is introduced stochastically (with probability
$0$ at the first iteration and probability $1$ at $t=\infty$) to
improve convergence. We will replace Eq.~\ref{eq:rules-bp2} with
Eqs.~\ref{eq:rules-rein3},\ref{eq:rules-rein4}:
\begin{eqnarray}
h_{i}^{t+1} & = & \frac{1}{\sqrt{n}}\sum_{b}\xi_{i}^{b}u_{b\rightarrow i}^{t}+\left\{ \begin{array}{ll}
0 & \textrm{w.p.}\,\gamma_{t}\\
h_{i}^{t} & \textrm{w.p.}\,1-\gamma_{t}\end{array}\right.\label{eq:rules-rein3}\\
h_{i\rightarrow a}^{t+1} & = & h_{i}^{t+1}-\frac{1}{\sqrt{n}}\xi_{i}^{a}u_{a\rightarrow i}^{t}\label{eq:rules-rein4}
\end{eqnarray}
We will use $\gamma_{t}=\gamma_{0}^{t}$ for $0\leq\gamma_{0}\leq1$
(though other choices are also possible). Choosing $\gamma_{0}=1$
clearly gives back the original BP set of equations, Eqs. \ref{eq:rules-bp1}-\ref{eq:rules-bp2}.
We note that a similar inertia term $\gamma h_i^t$ (constant $\gamma$) was introduced in~\cite{Murayama}, which would correspond to average the one in Eq. \ref{eq:rules-rein3}. Note also that the extra term for $\gamma_{t}=0$ corresponds to adding an
\emph{external field} equal to the local field computed in the last
step. Remembering that ``fixing'' a variable as in the standard
decimation procedure is equivalent to adding an external field of
infinite intensity, one can think of this procedure as a sort of \emph{smooth
decimation} in which \emph{all} variables (not only the most polarized
ones) get an external field, but the intensity is proportional to
their polarization.  Numerical experiments of learning randomly generated patterns have
been carried out on systems of various sizes (up to $n=10^{6}$),
with different choices of $K$ and with different topologies (overlapping
and tree--like). Some are reported in Fig.~\ref{fig:learning}. An
easy to use version of the code is made available at \cite{www-SP}.
It is not hard to think how the same algorithm could be made effective also in presence of faulty 
contacts and heterogeneous discrete synaptic values.
(which need not to be  identified a priori as the message-passing procedure, distributed over 
the same graph, could incorporate defects by modifying accordingly the messages).
 Even for the limit case of continuous synapses the process converge to optimal solutions in a
wide range of $\alpha$.

Experiments have been performed using an improved version of Eqs.~\ref{eq:rules-bp1}-\ref{eq:rules-bp2}:
Using further linearizations like in~\cite{Kabashima} one can obtain a new set of equations
that are equivalent to Eqs.~\ref{eq:rules-bp1}-\ref{eq:rules-bp2}
up to an error of $O\left(n^{-1/2}\right)$, having two main implementation
advantages: memory requirements of just $O\left(n\right)$ (in addition
to the set of patterns which amounts to $\alpha n^{2}$ bits), and
needing just $O\left(n\right)$ (slow) hyperbolic function computations
in addition to $O\left(n^{2}\right)$ elementary (fast) floating point
operations. 
BP equations can also be simplified by approximating $m_{k\rightarrow b}$
by $m_{k}$ in Eqs.~\ref{eq:rules-bp1}-\ref{eq:rules-bp2} (without correction terms), giving
a simple closed expression in the quantities $\{ m_{i}^{t}\}$. The
resulting equation is not asymptotically equivalent to BP anymore
(although the approximation itself has an error of $O\left(n^{-1/2}\right)$
it participates in a sum of $n$ terms), but nonetheless gives comparable
(just slightly worse) algorithmic performances. Of particular interest
are the corresponding equations for $\gamma_{0}=0$ (full reinforcement)
which take a simple additive form if written in terms of the local
fields $h_{i}^{t}$:
\begin{equation}
h_{i}^{t+1}=\sum_{t'\leq t}\sum_{b}\frac{\xi_{i}^{b}}{\sqrt{n}}u_{b}^{t}\,\,\sim\:\, h_{i}^{\tau+1}=h_{i}^{\tau}+\frac{\xi_{i}^{b}}{\sqrt{n}}u_{b_{\tau}}^{\tau}
\label{eq:almostseq}
\end{equation}
where $u_{b}^{s}=f\left(\sum_{k\neq i}\frac{\xi_{k}^{b}}{\sqrt{n}}\tanh h_{k}^{s},\frac{1}{n}\sum_{k\neq i}\tanh^{2}h_{k}^{s}\right)$
and $t$ scales as $\alpha n\tau$. By choosing at time $\tau$ one
pattern $\xi_{b_{\tau}}$ from the set $\Xi$, Eq.~\ref{eq:almostseq}
implements a sequential learning protocol, still leading to an extensive
memory capacity (around $\alpha_{max}\simeq.5$ for the binary perceptron).
The simplicity of Eq.~\ref{eq:almostseq} represents a proof-of-concept
of how highly non-trivial learning can take place by message-passing
between simple devices disposed over the network itself. This fact
could shed some light on the biological treatment of information in
neural systems~\cite{next}.

This work has been supported by the EC, MTR 2002-00319 'STIPCO' and FP6 IST consortium 'EVERGROW'. We thank the ISI Foundation for hospitality.

\end{document}